\documentclass{aa}
\usepackage{amsmath,graphics,amssymb}
\usepackage[varg]{txfonts}
\usepackage{natbib}
\usepackage{times}
\usepackage{frcursive}
\usepackage{color}

\usepackage[hidelinks]{hyperref}

\newfont{\eufont}{eufm10}

\newcommand{\zav}[1]{\left(#1\right)}
\newcommand{\hzav}[1]{\left[#1\right]}

\newcommand{\kms}{\ensuremath{\text{km}\,\text{s}^{-1}}}
\newcommand{\vel}{{v}}

\newcommand{\ms}{\ensuremath{{M}_{\odot}}}
\newcommand{\msr}{\ensuremath{\ms\,\text{yr}^{-1}}}
\newcommand{\ergs}{\ensuremath{\text{erg}\,\text{s}^{-1}}}

\newcommand{\de}{\mathrm{d}}
\newcommand{\Teff}{\mbox{$T_\mathrm{eff}$}}

\newcommand\x[1]{\ensuremath{#1_\text{X}}}
\newcommand\lx{\ensuremath{\x L}}
\newcommand\irr{\text{irrad}}

\newcommand\cc{\ensuremath{C_\text{c}}}

\newcommand{\Bsupergiant}{B\protect\nobreakdash-supergiant}

\begin{document}

\title{X-ray irradiation of the stellar wind in HMXBs with B~supergiants:
Implications for ULXs}

\author{J.~Krti\v{c}ka\inst{1} \and J.~Kub\'at\inst{2} \and
        I.~Krti\v ckov\'a\inst{1}}

\institute{\'Ustav teoretick\'e fyziky a astrofyziky,
           P\v{r}\'\i rodov\v edeck\'a fakulta, Masarykova univerzita,
           CZ-611 37 Brno, Czech Republic
           \and
           Astronomick\'y \'ustav, Akademie v\v{e}d \v{C}esk\'e
           republiky, CZ-251 65 Ond\v{r}ejov, Czech Republic}

\date{Received}

\abstract{Wind-fed high-mass X-ray binaries are powered by accretion of the
radiatively driven wind of the luminous component on the compact star. Accretion-generated X-rays alter the ionization state of the wind. Because higher
ionization states drive the wind less effectively, X-ray ionization may brake
acceleration of the wind. This causes a decrease in the wind terminal velocity and
mass flux in the direction toward the X-ray source. Here we study the effect of
X-ray ionization on the stellar wind of B supergiants. We determine the binary
parameters for which the X-ray irradiation significantly influences the stellar
wind. This can be conveniently studied in diagrams that plot the optical depth
parameter versus the X-ray luminosity. For low optical depths or for high X-ray
luminosities, X-ray ionization leads to a disruption in the wind aimed toward
the X-ray source. Observational parameters of high-mass X-ray binaries with
B-supergiant components appear outside the wind disruption zone. The X-ray
feedback determines the resulting X-ray luminosity. We recognize two states with
a different level of feedback. For low X-ray luminosities, ionization is weak, and
the wind is not disrupted by X-rays and flows at large velocities, consequently
the accretion rate is relatively low. On the other hand, for high X-ray
luminosities, the X-ray ionization disrupts the flow braking the acceleration,
the wind velocity is low, and the accretion rate becomes high. These effects
determine the X-ray luminosity of individual binaries. Accounting for the X-ray
feedback, estimated X-ray luminosities reasonably agree with observational
values. We study the effect of small-scale wind inhomogeneities (clumping),
showing that clumping weakens the effect of X-ray ionization by increasing
recombination and the mass-loss rate. This effect is particularly important in the
region of the so-called bistability jump. We show that ultraluminous X-ray
binaries with $\lx\lesssim10^{40}\,\ergs$ may be powered by accretion of a
B-supergiant wind on a massive black hole.}


\maketitle

\section{Introduction}

High-mass X-ray binaries (HMXBs) harbor a luminous massive star accompanied by
a degenerate object, either a neutron star or a black hole. In a class of these
binaries, the compact companion sails through the wind, which blows from the
massive star, powering the X-ray emission via wind accretion
\citep{davos,laheupet,velevela}.

Winds of hot stars are driven by light absorption and scattering in lines of
heavy elements such as carbon, silicon, and iron \citep{lusol,cak,ppk}. The
strength of the wind is typically characterized by its mass-loss rate and
terminal velocity. The mass-loss rate is defined as an amount of mass lost by
the star per unit of time and the terminal velocity is a limiting wind velocity
at large distances from the star. These parameters mostly determine the
influence of the wind on the stellar evolution and circumstellar medium and can
be estimated either from observations or theory.

The radiative force depends on the ionization state of the wind. Consequently,
the X-ray irradiation may change the ionization balance and significantly alter
the structure of the stellar wind. Because ions with a higher charge have
effectively a lower number of levels and their resonance lines are outside the
flux maximum, the X-ray irradiation weakens the radiative force. In HMXBs, this
leads to a reduction in the terminal velocity and possibly also in the mass flux
of the wind in a relatively narrow cone that faces the compact companion
\citep{sekerka,ff,velax1}.

The effect of X-ray irradiation provides important feedback for the wind
structure. The amount of accreted matter and therefore also the X-ray luminosity
is linearly proportional to the mass-loss rate, but inversely proportional to
about the fourth power of wind velocity. Therefore, a decrease in the wind terminal
velocity due to the X-ray irradiation significantly increases X-ray emission
\citep{hohoho,irchuch,sandvelax}.

Such feedback might be especially important for ultraluminous X-ray sources
(ULXs). These objects have X-ray luminosities higher than would correspond to the
Eddington limit of a typical stellar black hole \citep{atapka}. Therefore, they
were suspected to host intermediate mass black holes. While this still may be
true for many of these sources, the detection of X-ray pulsations in some of the
ULXs \citep{bachet,prvnidruhy} indicates that at least some of these sources may
be powered by accretion on a neutron star. The strong influence of X-rays on the
wind terminal velocity combined with high wind mass-loss rates could provide
an explanation for the enormous X-ray luminosity of these objects.

While the X-ray irradiation has been systematically studied in HMXBs with O star
primaries, such studies in the B star domain are only scarce
\citep[e.g.,][]{sandvelax}. However, the domain of B supergiants is particularly
interesting for HMXBs due to the bistability jump in mass-loss rates
and terminal velocities, which can particularly boost the X-ray luminosity
\citep{vinbisja}. Therefore, here, we provide a grid of hot star wind models of B
supergiants with X-ray irradiation focusing on the importance of the bistability
jump and its relevance to ULXs.

\section{Wind models}

\subsection{Global models without X-ray irradiation}

Wind modeling was based on the grid of {\Bsupergiant} METUJE models described in
detail by \citet{bcmfkont}. Our models were calculated assuming a spherically
symmetric and stationary stellar wind. The models self-consistently solve the
same equations in the photosphere and in the wind, which enables a smooth
transition from the photosphere to the wind (global models). The radiative
transfer was solved in the comoving frame \citep[CMF;][]{mikuh}. The atomic level
occupation numbers were determined from the kinetic equilibrium equations \citep
[abbreviated as NLTE,] [Chapter~9]{hubenymihalas} with radiative bound-free
terms calculated from the CMF radiative field and radiative bound-bound terms
with the Sobolev approximation \citep{klecany}. Atomic data for the solution of
kinetic equilibrium equations were adopted mostly from the TLUSTY models
\citep{bstar2006} with additional updates from the Opacity and Iron Project data
\citep{topt,zel0}. We assumed a solar chemical composition after \citet{asp09}.
The wind density, velocity, and temperature were derived from the continuity
equation, the equation of motion with a radiative force due to continuum and
line transitions, and the equation for energy \citep[see][for details]{kubii,kpp}.
We used the TLUSTY plane-parallel static model atmospheres
\citep{ostar2003,bstar2006} to derive the initial estimate of the photospheric
structure.

The models were calculated for a grid of effective temperatures $\Teff
=15\,000-25\,000\,$K, assuming fixed luminosity $L$ (and stellar mass $M$). The
list of adopted parameters given in Table~\ref{bvele} was further supplemented
by corresponding stellar radii $R_{*}$ and predicted mass-loss rates $\dot M$.
The predicted mass-loss rate of the model 150-60 is about five times higher than
for the models with higher effective temperatures. This is a consequence of
the so-called bistability effect \citep{bista}. This effect is caused by the
recombination of iron from \ion{Fe}{iv} to \ion{Fe}{iii}, which accelerates wind
more efficiently \citep{vikolabis,bcmfkont}.

\begin{table}[t]
\caption{Stellar parameters of the model grid with derived values of the
mass-loss rate $\dot M$ for smooth wind ($C_1=1$) and for winds with $C_1=10$
(see Eq.\,\eqref{najc}).}
\centering
\label{bvele}
\begin{tabular}{ccccc}
\hline
\hline
Model &$\Teff$ & $R_{*}$ & $\dot M(C_1=1)$ & $\dot M(C_1=10)$ \\
& $[\text{K}]$ & $[{R}_{\odot}]$ & \multicolumn{2}{c}{[\msr]} \\
\hline
\multicolumn{5}{c}{$M=60\,{M}_{\odot}$, $\log(L/L_\odot)=5.88$} \\
250-60 & 25000 & 46.5 & $3.2\times10^{-7}$ & $3.6\times10^{-7}$ \\
200-60 & 20000 & 72.7 & $4.3\times10^{-7}$ & $1.3\times10^{-6}$ \\
150-60 & 15000 & 129  & $2.1\times10^{-6}$ & $2.7\times10^{-6}$ \\
\hline
\end{tabular}
\end{table}

Hot star winds show a small-scale structure which most likely originates due to
the line-driven wind instability \citep{ocr,felto,sundsim}. The inhomogeneities, also referred to as clumping, soften the influence of X-ray irradiation due to
enhanced recombination \citep{osfek,irchuch}. To understand the influence of
clumping, we additionally calculated a set of models that account for small-scale
inhomogeneities \citep[described in][]{irchuch}. We assumed that the stellar
wind consists of homogeneous, optically thin overdensities (clumps) immersed in
void interclump space. We adopted a smooth velocity profile (describing the mean
flow) for our modeling because the numerical simulations of line-driven wind
instability predict that overdensities move at a velocity corresponding to
a stationary wind solution \citep{felpulpal,opsim,runow} and that they do not directly
influence the mass-loss rate. However, the line driving force in these numerical
simulations is calculated using fixed line force parameters, that is  different
from what we used in our calculations. Consequently, these simulations do not
account for the influence of overdensities on the level populations, which is what shall
be included using our NLTE models. Therefore, within our assumptions, clumping
affects just free-bound and free-free processes whose rates scale with the square of
the density \citep[see][for details]{irchuch}. This is also a standard approach
for spectral analysis of clumped winds \citep{hamko,hilmiwc,pulchuch}.

The models with clumping are parameterized by a clumping factor
$\cc={\langle\rho^2\rangle}/{\langle\rho\rangle^2}$, where the angle brackets
denote the average over volume. The clumping factor describes the density of the
clump relatively to the mean density. We adopted radial clumping stratification
motivated by empirical studies \citep{najradchuch,bouhil}
\begin{equation}
\label{najc}
\cc(r)=C_1+(1-C_1) \, e^{-{\vel(r)}/{C_2}},
\end{equation}
which grows from unity in the photosphere (this is what corresponds to a smooth wind) to
$C_1$ for velocities larger than $C_2$. We adopted the same values as in
\citet{bcmfkont}, that is $C_1=10$, which is close to the mean value for which
the empirical H$\alpha$ mass-loss rates of B supergiants agree with observations
\citep{bcmfkont} and $C_2=100\,\kms$, which is a typical value derived in the
observational study of \citet{najradchuch}. In the formula for the radial
clumping stratification Eq.~\eqref{najc}, we inserted the fit $\tilde \vel (r)$
of the velocity of the smooth wind model ($\cc=1$) via a
modified polynomial form from \citet{betyna} of
\begin{equation}
\label{vrfit}
\tilde \vel (r)=\sum_i \varv_i\zav{1-\gamma\frac{R_*}{r}}^i,
\end{equation}
where $\varv_i$ and $\gamma$ are parameters of the fit given in Table\,2 from
\citet{bcmfkont}.

The mass-loss rates predicted from models with clumping are given in the last
column of Table~\ref{bvele}. Clumping causes stronger recombination, which
lowers the ionization state of atoms. Since lower ionization states typically
accelerate the wind more effectively, clumping leads to a higher mass-loss rate
\citep{muij,irchuch}. This effect is the strongest for the model 200-60, where
the clumping causes an earlier onset for the bistability effect.

\subsection{Inclusion of X-ray irradiation into NLTE models}
\label{kap_inx}

The presence of an external source of X-ray radiation breaks the large-scale spherical
symmetry of the stellar wind. The flow is disrupted by the gravity of the
compact object and the accretion wake trailing the compact
companion forms \citep{blondyn,manousci}.
Moreover, as a result of the weakening of the radiative force by X-rays, an
ionization wake may form \citep{ff,felabon}. On small scales, accretion of
clumped wind contributes to X-ray variability \citep{osfek,bozof,elmel}.

Self-consistent modeling of such time-dependent phenomena requires
multidimensional hydrodynamical simulations. However, coupling these
simulations with a solution for the radiative transfer equation together with
determination of atomic level population (i.e., the NLTE problem) is
computationally prohibitive, and no such models are currently available. To make
the problem tractable, we solved the stationary hydrodynamical equations
assuming that the derived solution describes properties of the mean flow. We
only accounted for the radial motion of the fluid, which is expected to be
dominant in most cases. These approximations allow us to understand the
influence of X-rays on the radiative force using 1D models, while the detailed
3D structure of the flow should be derived from hydrodynamical simulations.
A similar approach was also used by other authors \citep{sandvelax}.

Similarly to \citet{irchuch}, the influence of the compact secondary is only
taken into account by the inclusion of external X-ray irradiation.
We assume a point irradiating source located at the distance $d$ from a
given point in the wind, in which case the term
\begin{equation}
\label{xneutron}
J_\nu^\text{X}=\frac{L_\nu^\text{X}}{16\pi^2d^2}\text{e}^{-\tau_\nu(r)}
\end{equation}
has to be added to the mean radiation intensity $J_\nu$. Here, $L_\nu^\text{X}$
is the monochromatic X-ray irradiation luminosity, whose frequency dependence is
approximated by the power law $L_\nu^\text{X}\sim\nu^ {-\Gamma}$ for energies
from 0.5 to 20~keV with a power law index $\Gamma=1$. The monochromatic X-ray
irradiation luminosity is normalized by the total X-ray luminosity, $\x L=\int
L_\nu^\text{X}\,\de\nu$, which enters our models as a free parameter. A second
free parameter of our models is the binary separation $D$, which is the distance
between stellar centers \citep[see Fig.\,2 in][]{irchuch}. The binary separation
enters the expression for the distance of a given point from the compact
companion. Finally, $\tau_\nu(r)$ is the frequency-dependent optical depth
between the given point in the wind and the compact companion,
\begin{equation}
\label{zamlhou}
\tau_\nu(r)=\left|\int_r^D\kappa_\nu(r')\rho(r')\,\de r'\right|,
\end{equation}
where $\kappa_\nu$ is the X-ray mass-absorption coefficient.

Eq.~\eqref{xneutron} allows one to calculate wind models for different
inclinations with respect to the binary axis. Such models show that the
influence of the X-ray irradiation is largest for zero inclination, that is,
along the ray connecting stellar centers \citep{velax1}. Therefore, we
calculated wind models only along the direction of the binary axis.

\begin{table}[t]
\caption{Coefficients of the fit of the averaged mass-absorption coefficient
used in Eq.~\eqref{kapafit}.}
\label{fitko}
\centering
\begin{tabular}{rcccccc}
\hline
\hline
$C_1$ & $\lambda_1$ & $a_0$ & $a_1$ & $b_1$ & $a_2$ & $b_2$\\
\hline
\multicolumn{7}{c}{Model 250-60}\\
$1$ & 20.1799 & 220 & 2.514 & $-0.784$ & 2.641 & $-1.440$\\
$10$& 20.1799 & 210 & 2.504 & $-0.774$ & 2.371 & $-1.004$\\
\hline
\multicolumn{7}{c}{Model 200-60}\\
$1$ & 20.1799 & 195 & 2.420 & $-0.741$ & 2.642 & $-1.441$\\
$10$& 23      &     & 1.594 & $0.488$ & 2.028 & $-0.234$\\
\hline
\multicolumn{7}{c}{Model 150-60}\\
$1$ &  23     &     & 1.401 & $1.075$ & 1.644 & $0.678$\\
$10$&  23     &     & 1.336 & $1.452$ & 1.425 & $1.300$\\
\hline
\end{tabular}
\end{table}

\begin{figure*}[t]
\centering
\resizebox{0.49\hsize}{!}{\includegraphics{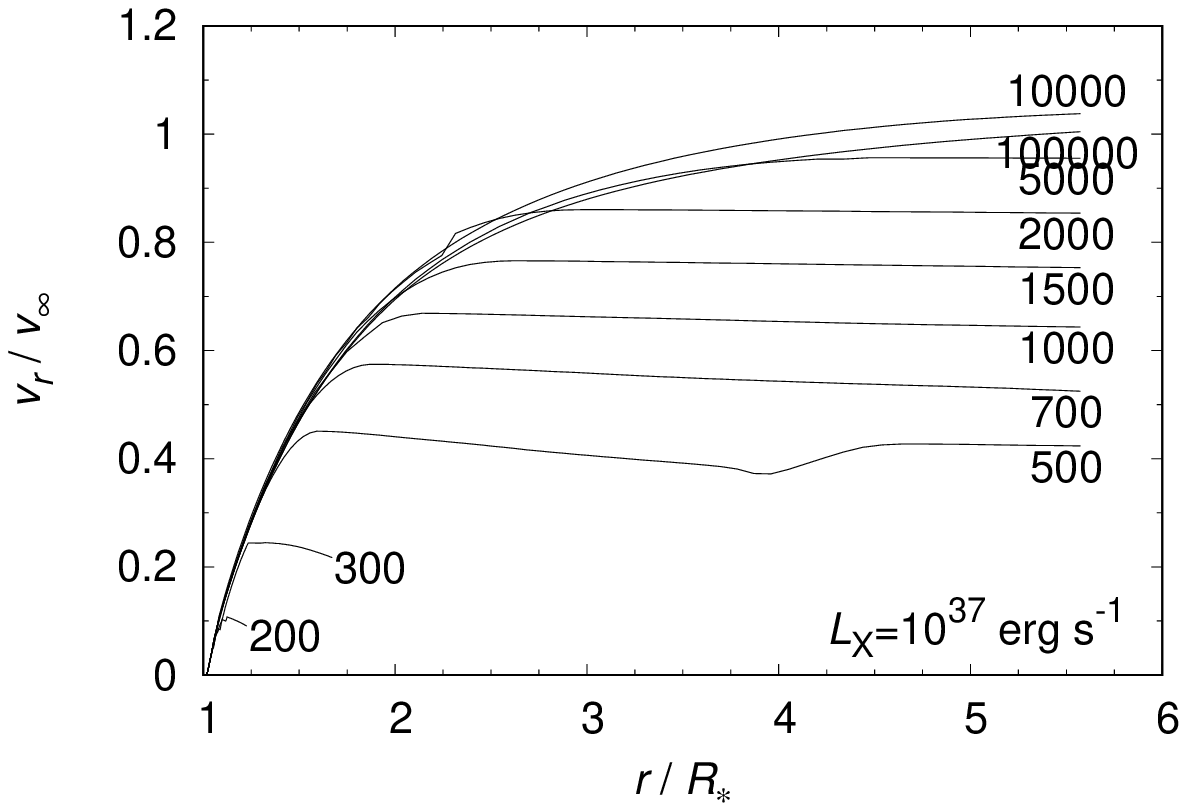}}
\resizebox{0.49\hsize}{!}{\includegraphics{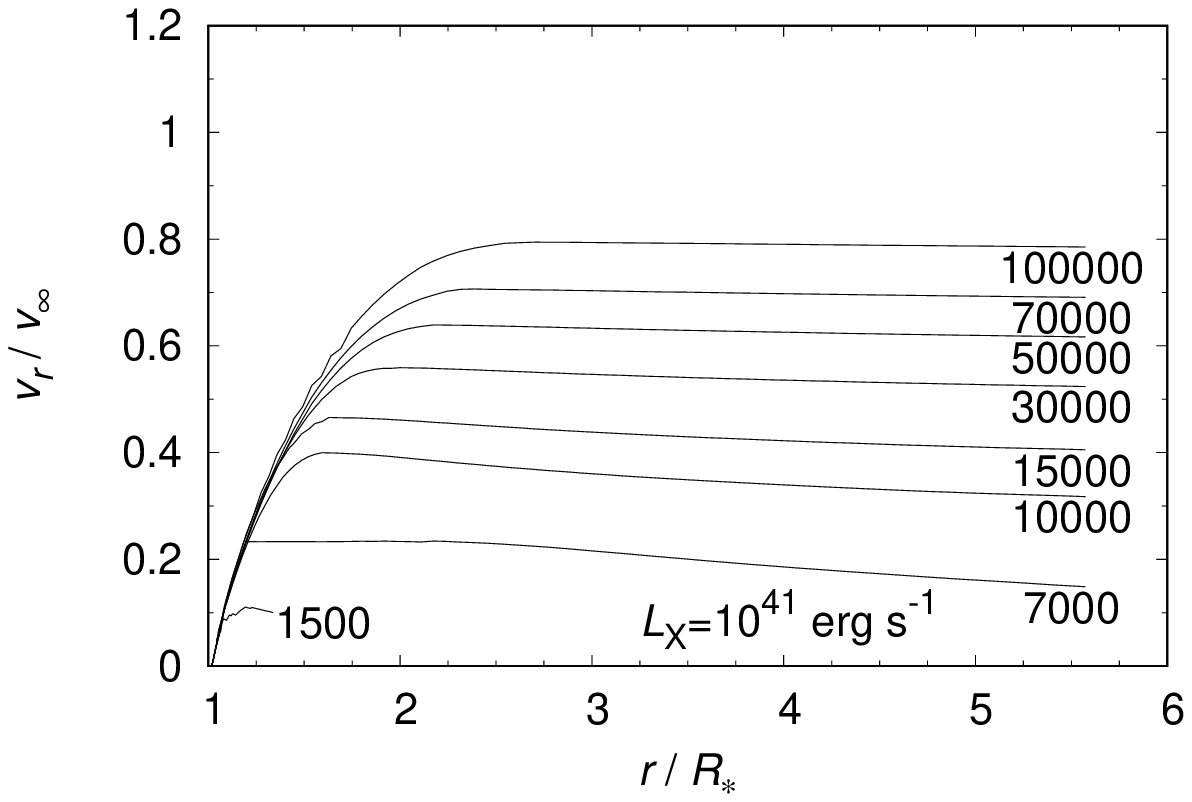}}
\caption{Radial variations of velocity in the model 150-60 without clumping with
the total X-ray luminosity
$\x L=10^{37}\,\ergs$ ({\em left panel}) and $\x L=10^{41}\,\ergs$ ({\em right
panel}). Individual curves are labeled by a binary separation $D$ in units of
${R}_{\odot}$.}
\label{vr15060}
\end{figure*}
\begin{figure*}[t]
\centering
\resizebox{0.49\hsize}{!}{\includegraphics{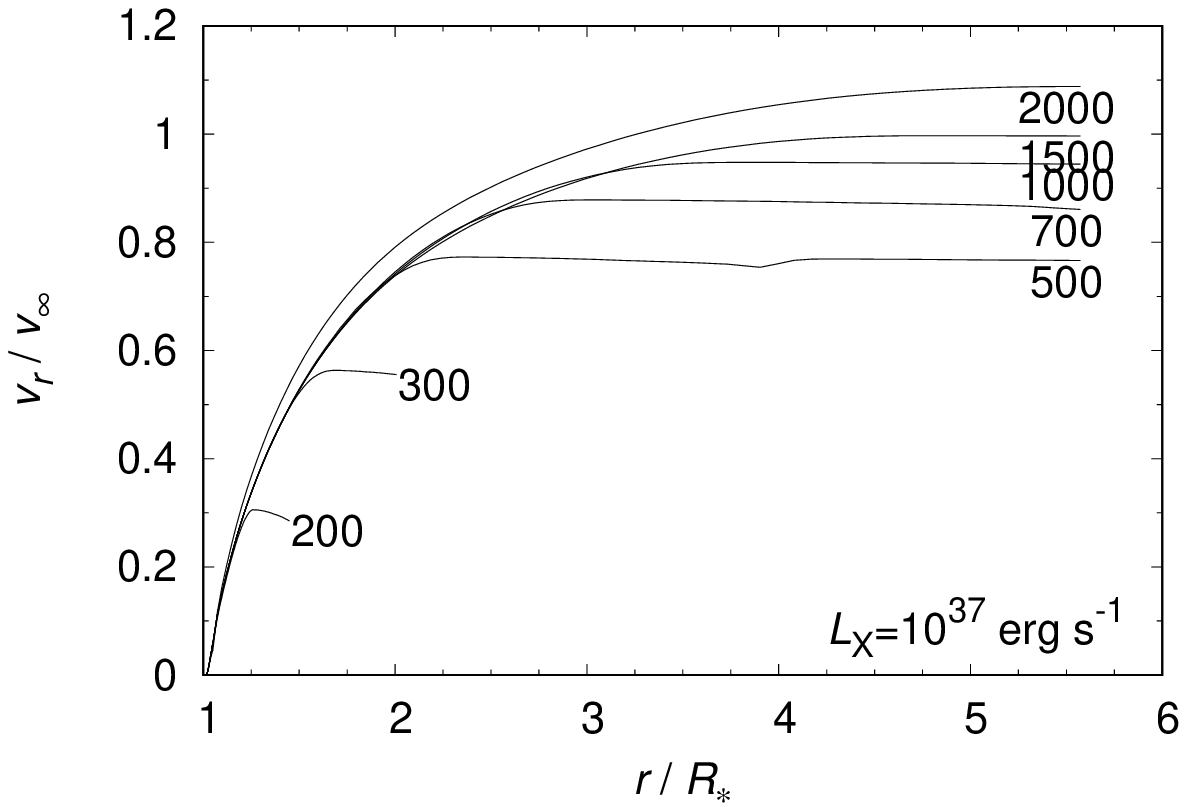}}
\resizebox{0.49\hsize}{!}{\includegraphics{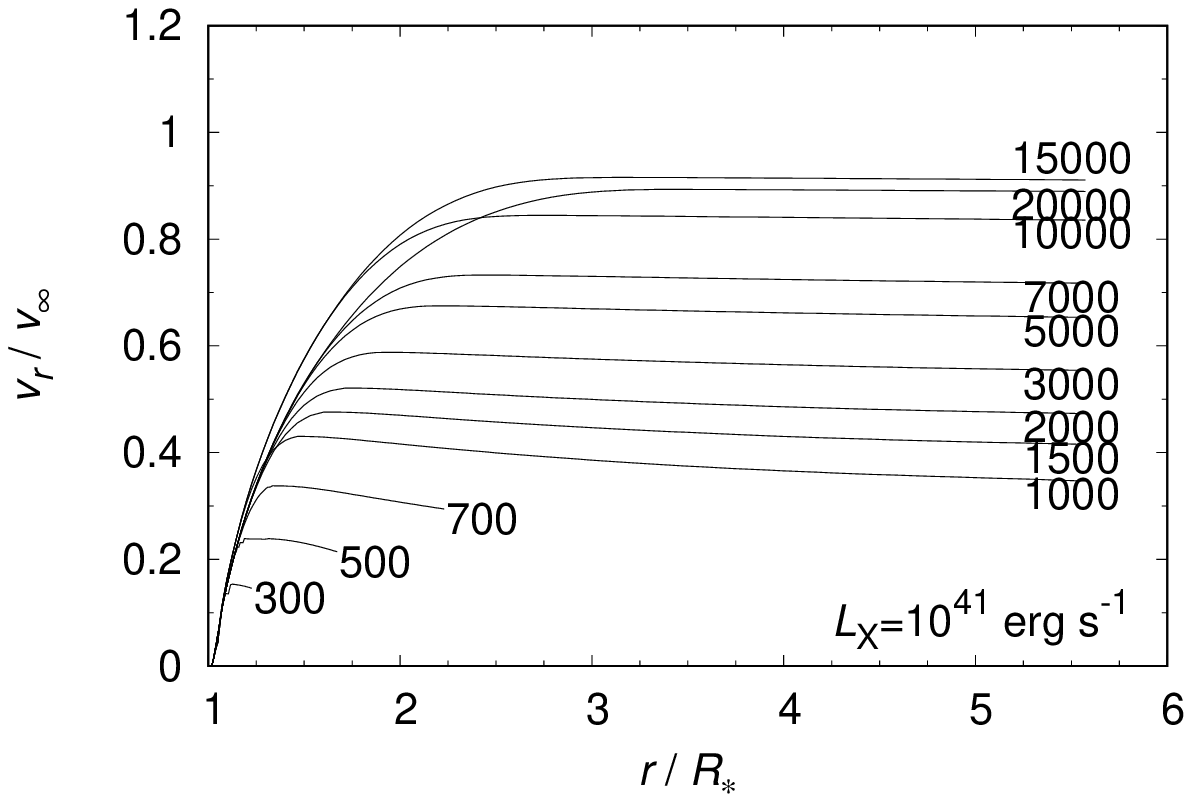}}
\caption{Same as Fig.~\ref{vr15060}, but for clumping factor $ C_1 =10$.}
\label{vr15060ch10}
\end{figure*}

To avoid possible problems with the convergence of the models, we did not use the
density and opacity from the actual model in Eq.~\eqref{zamlhou}. Instead, we
used the following analytical formula: 
\begin{equation}
\label{anhydrid}
\begin{aligned}
\rho(r)&=\frac{\dot M}{4\pi r^2 v(r)},\\
\varv(r)&=\min(\tilde \varv(r),\varv_\text{kink}),\\
\kappa_\nu(r)&=\tilde \kappa_\nu^\text{X}.
\end{aligned}
\end{equation}
Here $\varv_\text{kink}$ is the velocity of the kink that appears in the models
with strong irradiation (otherwise we put $\varv_\text{kink}\rightarrow\infty$),
$\tilde \varv(r)$ is the fit from Eq.~\eqref{vrfit} of the wind velocity derived
from the models without X-ray irradiation, and $\tilde \kappa_\nu^\text{X}$ is
the radially averaged mass-absorption coefficient given by\footnote{Within this
work, $\log$ stands for the decadic logarithm.}
\begin{equation}
\label{kapafit}
\log\zav{\frac{\tilde \kappa_\nu^\text{X}}{1\,\text{cm}^2\,\text{g}^{-1}}}=
  \left\{\begin{array}{l}
    \min(a_1\log\lambda+b_1,\log a_0),
    \quad \lambda<\lambda_1,\\
    a_2\log\lambda+b_2,\quad \lambda>\lambda_1,\\
  \end{array}\right.\\
\end{equation}
where $\lambda$ is the value of the wavelength in units of \AA. The parameters
$\lambda_1$, $a_0$, $a_1$, $b_1$, $a_2$, and $b_2$ given in Table~\ref{fitko}
were determined by fitting the mass-absorption coefficient of the models without
X-ray irradiation averaged over radii $1.5\,R_*-5\,R_*$. From Table~\ref{fitko}
it follows that the parameters of the fit do not significantly vary with
clumping in most cases; consequently, clumping does not strongly alter the
opacity in the X-ray energy domain \citep{lojza,irchuch}. 

To avoid numerical instabilities and problems with the CMF radiative force solution
in the  presence of a nonmonotonic velocity law, we used the photospheric flux to
calculate the radiative force and applied a Sobolev line force corrected for CMF
radiative transfer \citep[see][]{velax1}. This approach slightly shifts the
stellar radius, which corresponds to the lower boundary of our models,
affecting the velocity law used to determine clumping stratification. To
compensate for this, we selected $\gamma$ in Eq.~\eqref{vrfit} in
such a way that it leads
to the same mass-loss rate as calculated by \citet{bcmfkont} for global models
with clumping (these mass-loss rates are also given in Table~\ref{bvele}).

\section{Influence of the X-ray irradiation on the wind structure of B
supergiants}

X-ray irradiation leads to stronger ionization, which means that the
fraction of ions with higher ionization energies becomes higher. For weak X-ray
irradiation, this leads to a slight increase in the radiative force because new
states that contribute to the radiative force appear and ions with lower
ionization energies remain nearly unaffected. However, strong X-ray irradiation
depopulates ions with lower ionization energies, which are significant
contributors to the radiative force. This leads to a decrease in the radiative
force \citep{irchuch,sandvelax}.

The abovementioned influence of X-ray irradiation is proportional to the
irradiating luminosity, which is inversely proportional to the electron number density
$n$ as a result of an increase in the recombination with density, and inversely
proportional to the square of distance from the X-ray source due to the spatial
dilution of radiation. This motivates the basic form of the ionization parameter
introduced by \citet[][see also \citealt{sekerka}]{tenci}. Moreover, the
recombination is stronger in clumped media \citep{osfek} and a part of the
emitted X-rays may be absorbed in the intervening media \citep{karbim}.
Consequently, \citet{irchuch} introduced the ionization parameter as
\begin{equation}
\label{xic}
\xi(r)=\frac{1}{n d^2C_\text{c}}
\int L_\nu^\text{X}\text{e}^{-\tau_\nu(r)}\,\de\nu.
\end{equation}
From this equation, it follows that the X-ray irradiation is especially important
in a close neighborhood of the X-ray source and for the X-ray sources with
higher luminosity.

The influence of the X-ray irradiation is demonstrated in Fig.~\ref{vr15060},
where we plotted radial variations of velocity of the model 150\nobreakdash -60
without clumping for two luminosities of the external irradiating source and
different locations of the source. Close to the star, the wind density is very
high and the dependence on the optical depth dominates Eq.~\eqref{xic}.
Consequently, the ionization parameter is very low and the wind velocity
corresponds to the case without X-ray irradiation. This changes in a close
proximity of the X-ray source, where the optical depth becomes lower than one
and other dependencies prevail in Eq.~\eqref{xic}.

Strong X-ray ionization leads to a decrease in the radiative force, which is
unable to accelerate the wind any more and to sustain the flow with
monotonically increasing velocity at a given mass flux \citep{feslop,fero}. As a
consequence, a kink in the radial dependence of velocity appears (see
Fig.~\ref{vr15060}). The position of the kink therefore marks the region with
a strong interaction of irradiating X-rays with the supergiant wind.
Fig.~\ref{vr15060} shows that with decreasing binary separation, the position of
the kink moves toward the star reflecting a stronger influence of X-rays on the
flow. The models with the shortest binary separations are not extended up to
large radii due to convergence problems that appear when the velocity kink is
located at low speeds. Moreover, the wind may not reach the companion in such
a case and it may fall back to the star.

When the X-rays start to influence the structure of the flow close to the point
where the wind velocity is equal to the Abbott speed \citep[of the radiative
acoustic waves,][]{abbvln} and where the wind mass-loss rate is determined, the
wind becomes inhibited by X-rays leading to a significant decrease in the wind
mass flux \citep{irchuch}. Therefore, for a given X-ray luminosity, there is a
minimum binary separation that does not lead to the wind inhibition. Wind
inhibition appears for binary separations that are lower than those plotted in
Fig.~\ref{vr15060}.

With higher irradiating luminosity, the influence of X-rays becomes stronger
(Fig.~\ref{vr15060}, right panel) and the position of the kink moves toward the
star. Therefore, with increasing X-ray luminosity, the minimum binary separation
for which the wind is not inhibited increases. The kink in $\varv(r)$ appears
even for extreme binary separations of about a hundred stellar radii for an X-ray
luminosity corresponding to the ULX regime ($\x L=10^{41}\,\ergs$).

Clumping reduces the influence of X-rays (Fig.~\ref{vr15060ch10}). With
clumping, recombination becomes more efficient, and therefore a closer or stronger
X-ray source is needed to disrupt the wind. Moreover, clumping leads to an increase
in the mass-loss rate, which further weakens the effect of X-rays. The reduction of
the influence of X-rays due to clumping appears in all model stars studied here and
becomes especially apparent for the model 150-60 with the highest mass-loss
rate. Here, only a model for a very close X-ray source of $D=200\,R_\odot$ and
extreme X-ray irradiation with $\lx=10^{41}\,\ergs$ leads to wind inhibition.

We additionally calculated a small set of models with a modified index of spectral
energy distribution of irradiating X-rays $\Gamma=1.5$ to test the influence of
this parameter. The results showed that the radius where the kink of the
velocity profile appears is typically shifted by less than a few percent.
Therefore the slope of irradiating X-ray emission does not significantly
influence the final results.

We have described the influence of X-rays on the wind terminal velocity and mass-loss
rate using similar parametric relations as \citet{irchuch}. The terminal
velocity in the direction of the companion as a function of X-ray luminosity and
binary separation was expressed for the models with clumping using modified
equation (14) from \cite{irchuch} as
\begin{multline}
\label{ch10najventaurov}
v_\infty(\lx,D)=v_{\infty,0}\zav{1-\frac{R_*}{D}}^{\beta_1(\lx/L_{36})^{\beta_2}}\\*
+\Delta v_{\infty}\text{e}^{-(\log(D/R_*)-d_1\log(\lx/L_{36})-d_2)^2/d_3}.
\end{multline}
Here $L_{36}=10^{36}\,\ergs$ and the values of fit parameters $v_{\infty,0}$,
$\beta_1$, $\beta_2$, $\Delta v_{\infty}$, $d_1$, $d_2$, and $d_3$, which were
determined from the fit of the results from our models, are given in
Table~\ref{ch10najventautabv}. The first term in Eq.~\eqref{ch10najventaurov}
describes the increase in the terminal velocity with a decreasing influence of X-rays,
while the exponential term accounts for a peak in the terminal velocity for medium
irradiation. The dependence of the predicted mass fluxes on the X-ray luminosity
and binary separation, which roughly describes the effect of wind inhibition,
can be approximated as
\begin{equation}
\label{ch10najventaurovdmdt}
\dot m(\lx,D)=\frac{\dot M(\lx,D)}{4\pi R_\ast^2}=\frac{\dot M_0}{4\pi R_\ast^2}
\hzav{1-\exp\zav{-\frac{\zav{D/R_*-1}^2}{s_1(\lx/L_{36})^{s_2}}}}.
\end{equation}
Here $\dot M_0$, $s_1$, and $s_2$ are parameters, which were determined by
fitting predicted mass-loss rates (see Table~\ref{ch10najventaurovdmdttab}). We
note that the constant $s_1$ is dimensionless here, while in \citet [Eq.~(15)]
{irchuch}, it is expressed in units of $R_\ast^2$. The region of parameters where
Eq.~\eqref{ch10najventaurovdmdt} predicts $\dot m(\lx,D)\ll \dot M_0/(4\pi
R_\ast^2)$ corresponds to wind inhibition. X-rays disrupt the flow only in parts
of the supergiant irradiated by the compact companion \citep{velax1} with a peak
at the line connecting stellar centers; therefore, the total mass-loss rate,
which can be derived by integration of mass fluxes over the supergiant surface,
is only slightly affected by irradiation.

\begin{table}[t]
\caption{Derived parameters of the terminal velocity fit in
Eq.~\eqref{ch10najventaurov} for individual models with X-ray irradiation and
clumping.}
\label{ch10najventautabv}
\centering
\begin{tabular}{cc@{\hspace{3mm}}c@{\hspace{3mm}}ccccc}
\hline
\hline
Model & $v_{\infty,0}$ & $\beta_1$ & $\beta_2$ & $\Delta v_{\infty}$ &
$d_1$ & $d_2$ & $d_3$\\
\hline
250-60 & 1.08 & 1.78 & 0.440 & 0.65 & 0.470 & 1.23 & 0.23\\
200-60 & 1.04 & 1.82 & 0.384 & 0\\
150-60 & 1.05 & 0.724 & 0.214 & 0\\
\hline
\end{tabular}
\tablefoot{Parameters $v_{\infty,0}$ and $\Delta v_{\infty}$ are expressed in
units of the terminal velocity without X-ray irradiation.
Missing parameter values were not applied.}
\end{table}

\begin{table}[t]
\caption{Derived parameters of the mass-loss rate fit in
Eq.~\eqref{ch10najventaurovdmdt} for individual models with X-ray irradiation
and clumping.}
\label{ch10najventaurovdmdttab}
\centering
\begin{tabular}{ccccc}
\hline
\hline
Model & $\dot M_0$ [\msr] & $s_1$ & $s_2$ \\
\hline
250-60 & $4.5\times10^{-7}$ & 0.085 & 0.843 \\
200-60 & $1.4\times10^{-6}$ & 0.032 & 0.624 \\
150-60 & $2.6\times10^{-6}$ & $2.9\times10^{-4}$ & 0.720 \\
\hline
\end{tabular}
\end{table}

\begin{figure*}
\includegraphics[width=0.33\textwidth]{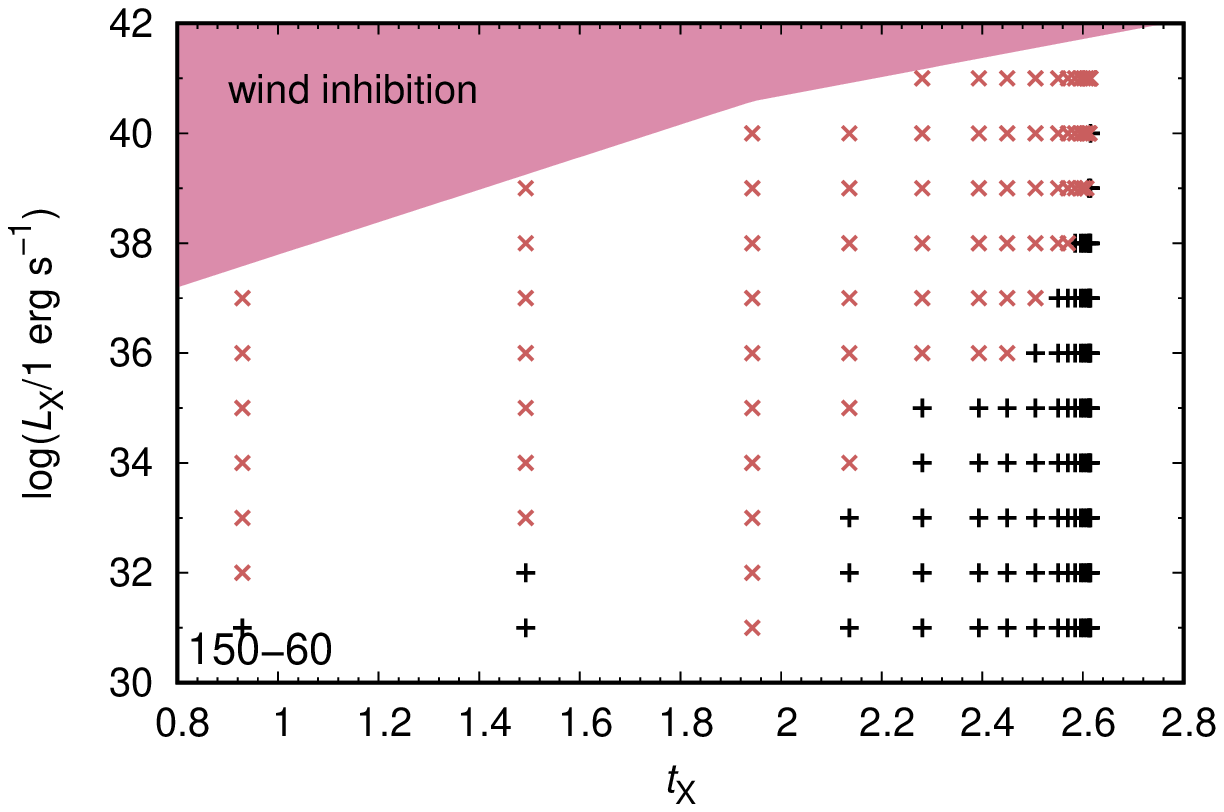}
\includegraphics[width=0.33\textwidth]{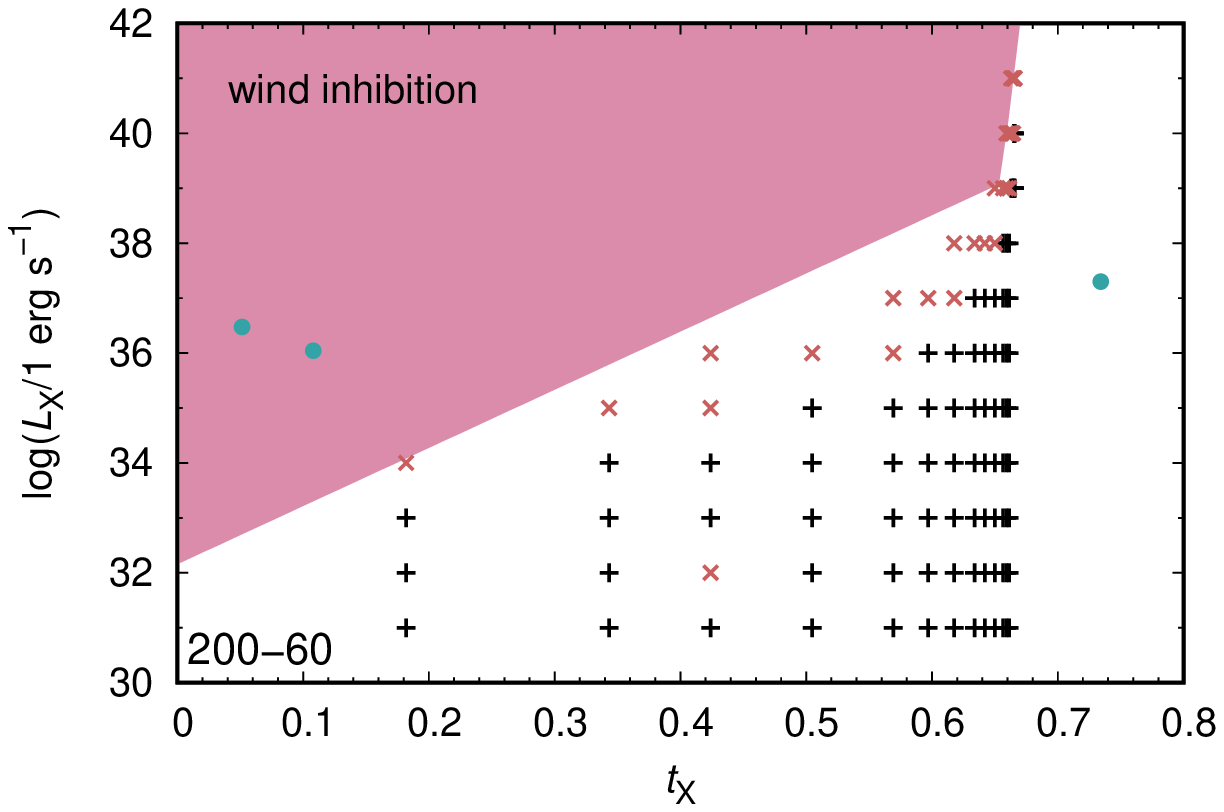}
\includegraphics[width=0.33\textwidth]{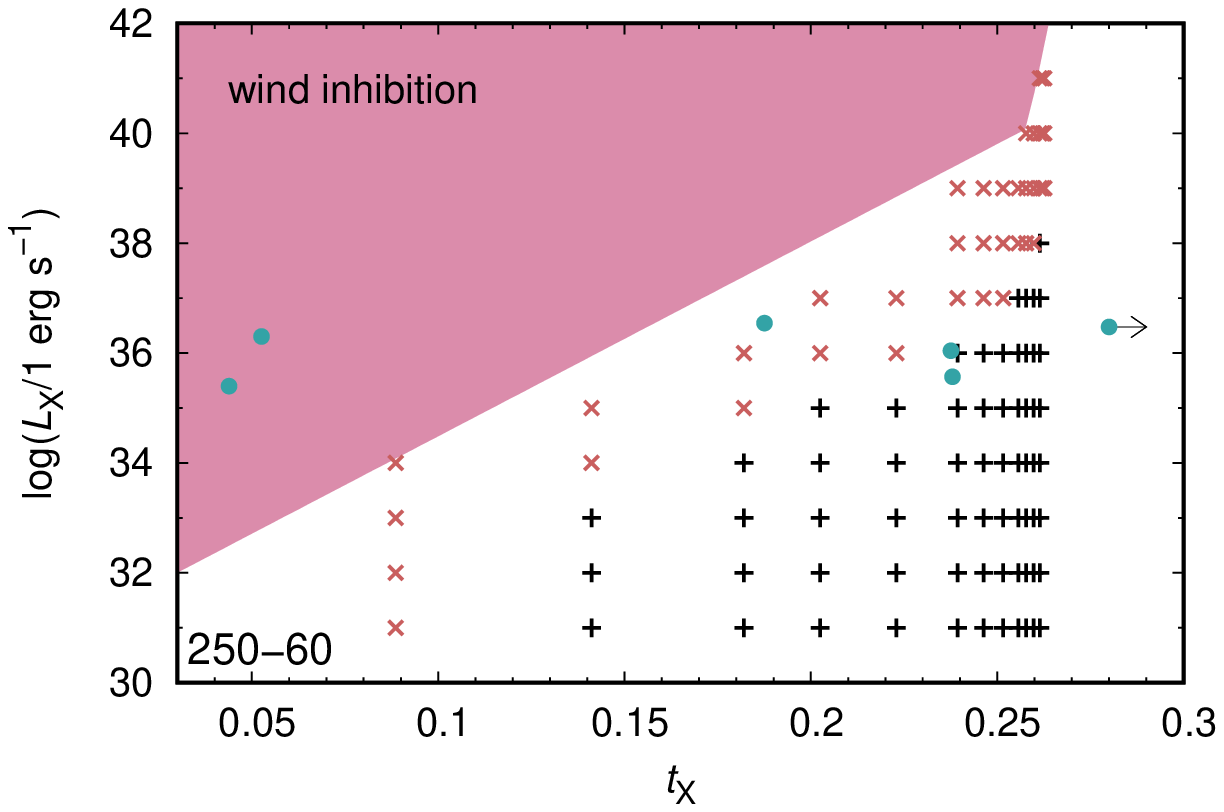}
\centering
\caption{Diagrams of the X-ray luminosity versus the optical depth parameter in
Eq.~\eqref{tx} for wind models without clumping of individual B supergiants with
parameters given in Table~\ref{bvele}. Individual symbols correspond to models
with different $\x L$ and $D$. Different symbols distinguish between various
effects of X-ray ionization on the wind: black plus symbols denote models with
a negligible influence of X-ray irradiation and red crosses denote models where
the X-ray irradiation reduces the terminal velocity. The antique pink area marks
the parameter region where the wind is inhibited by X-rays. The
positions of B supergiant components of HMXBs from Table~\ref{neutron} are overplotted (green
filled circles).}
\label{xiobr}
\end{figure*}

\begin{figure*}
\includegraphics[width=0.33\textwidth]{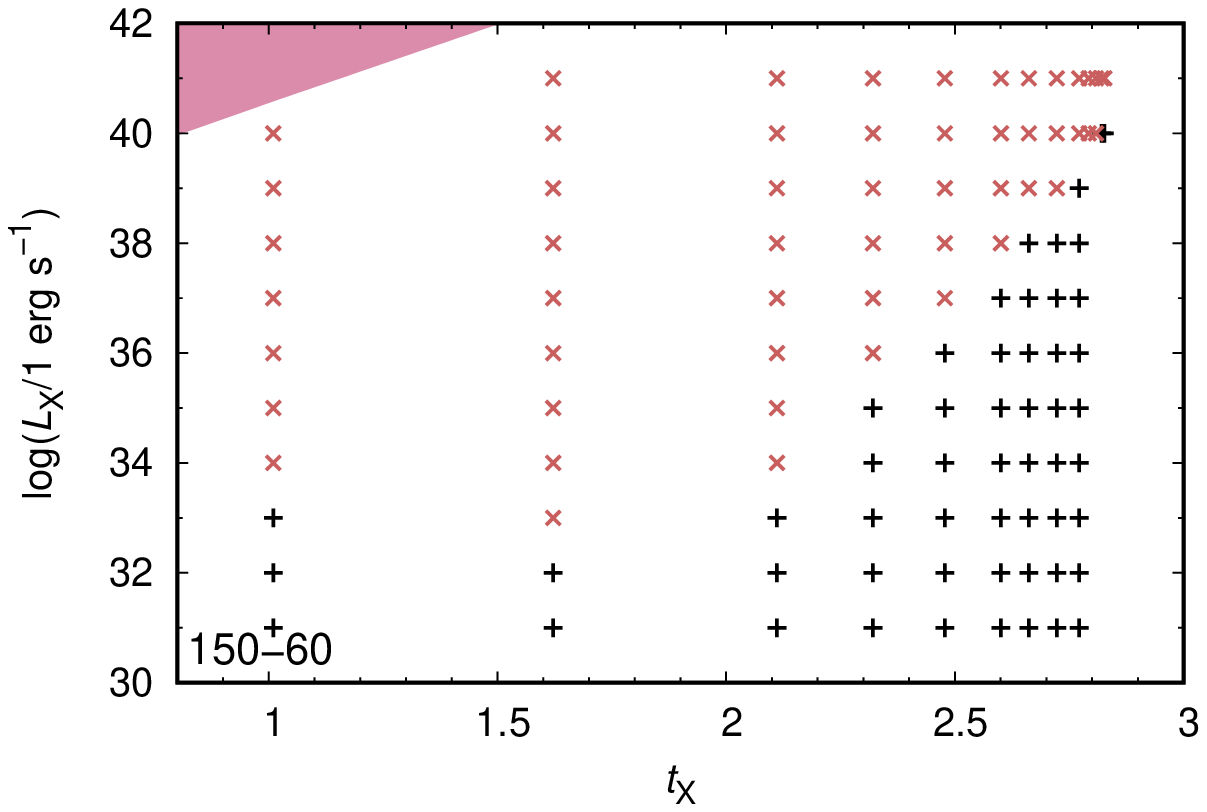}
\includegraphics[width=0.33\textwidth]{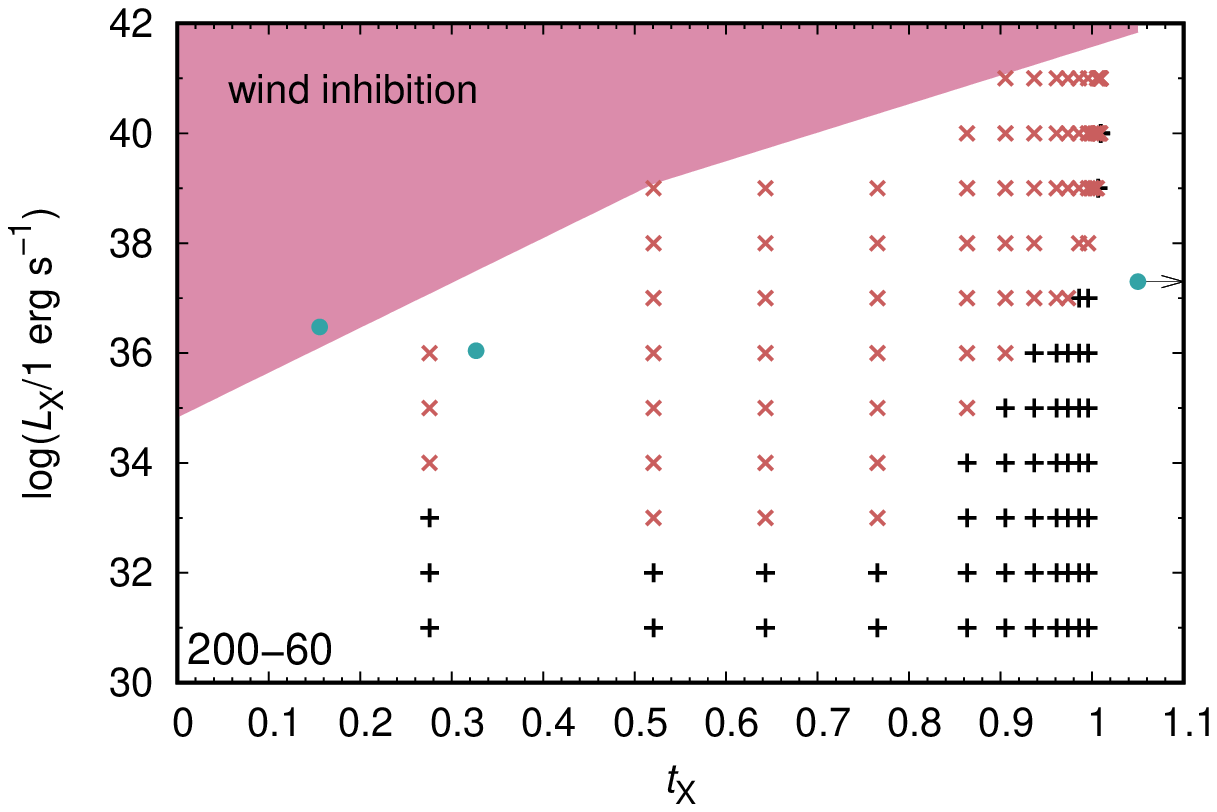}
\includegraphics[width=0.33\textwidth]{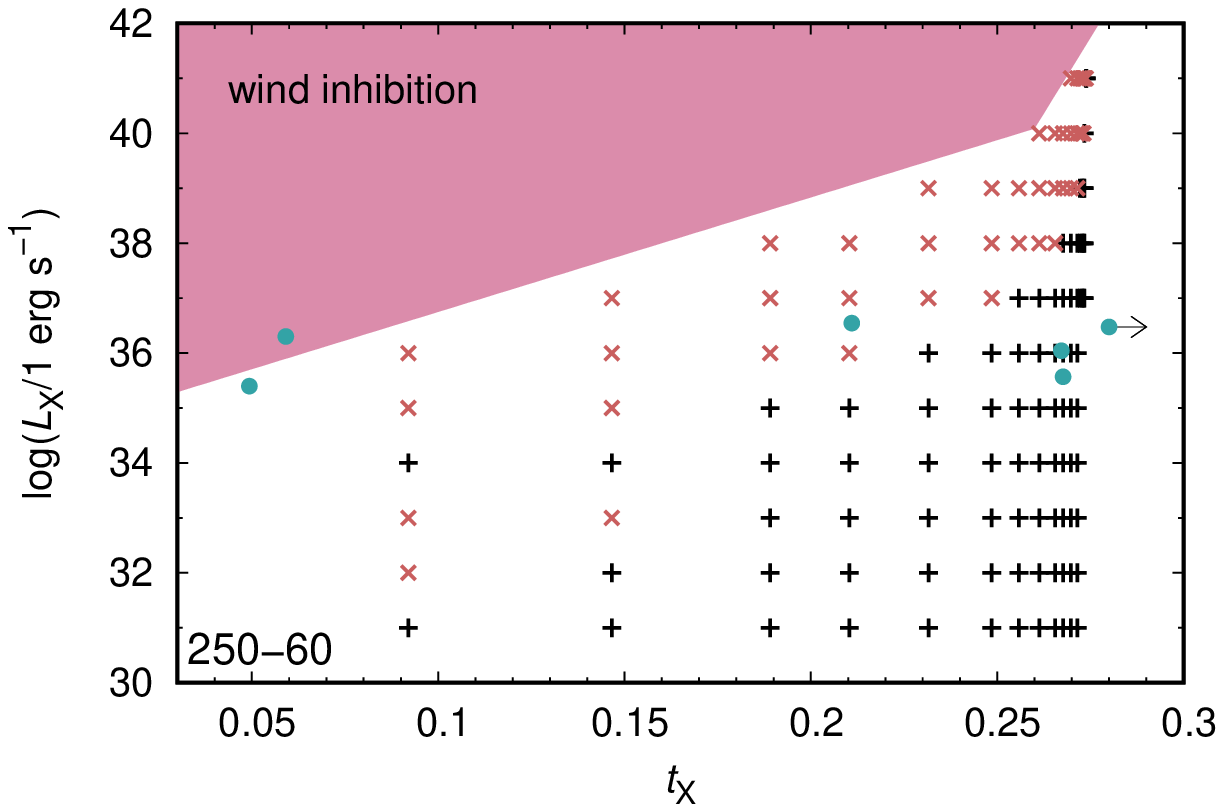}
\centering
\caption{As in Fig.~\ref{xiobr}, but for models with clumping.}
\label{xiobrch}
\end{figure*}

\section{Test against observations: Diagrams of the X-ray luminosity versus the
optical depth parameter}

In optically thick media, the influence of X-ray irradiation, which can be
quantified by the ionization parameter \eqref{xic}, mostly depends on the
optical depth between a given point in the wind and the X-ray source. In inserting
the density and opacity from Eq.~\eqref{anhydrid} into the expression for the
optical depth Eq.~\eqref{zamlhou} assuming that the wind has reached the
terminal velocity, the expression can be integrated to give $\tau_\nu(r)=\tilde
\kappa_\nu^\text{X} \dot M/(4\pi v_\infty)\zav{1/r-1/D}$. This motivated us to
introduce the optical depth parameter \citep [Eq.\,(3)] {dvojvit}
\begin{equation}
\label{tx}
\x t=\frac{\dot M}{\varv_\infty}\zav{\frac{1}{R_*}-\frac{1}{D}}
\zav{\frac{10^3\,\text{km}\,\text{s}^{-1}\,1\,R_\odot}
{10^{-8}\,{M}_\odot\,\text{yr}^{-1}}}
,\end{equation}
which is proportional to the radial optical depth between the stellar surface
and the X-ray source.

The diagrams that display the X-ray luminosity versus the optical depth
parameter were proven to be effective in separating domains according to the
type of influence that X-rays have on the wind \citep{irchuch}. These diagrams
are plotted for studied B supergiants in Figs.~\ref{xiobr} and \ref{xiobrch} for
the models without clumping and with clumping, respectively. From the diagrams,
it follows that the stellar wind is strongly influenced by X-ray irradiation,
either for high X-ray luminosities or for low X-ray optical depth parameters.
The latter in fact implies X-ray sources that are very close to the supergiant.
There is a zone where the X-ray irradiation becomes so strong that it is able
to inhibit the wind. No wind-fed X-ray binary should exist in the zone of wind
inhibition.

To test this prediction, we collected parameters of HMXBs with {\Bsupergiant}
primaries from the literature (see Table~\ref{neutron}) and placed these binary
systems into X-ray luminosity versus optical depth diagrams in Fig.~\ref{xiobr}.
Contrary to our prediction, a significant fraction of binaries lies in the
region where we expect wind inhibition.

\begin{table*}[t]
\caption{Parameters of HMXBs that appear in Figs.~\ref{xiobr} and \ref{xiobrch}.}
\label{neutron}
\centering
\begin{tabular}{l@{\hspace{2.5mm}}c@{\hspace{2.5mm}}c@{\hspace{2.5mm}}l@{\hspace{2.5mm}}c@{\hspace{2.5mm}}c@{\hspace{2.5mm}}c@{\hspace{2.5mm}}c@{\hspace{2.5mm}}c@{\hspace{2mm}}c@{\hspace{2.5mm}}c}
\hline
Binary & Sp. Type & $\log(L/L_\odot)$ & $T_\text{eff}$ [K] & $R_*$ [$R_\odot$] &
$M$ [$M_\odot$] & $D$ [$R_\odot$] & $\x L$ [$\text{erg}\,\text{s}^{-1}$]&
$\dot M$ [$M_\odot\,\text{yr}^{-1}$] & \x t & Reference\\
\hline
\object{IGR J00370+6122\tablefootmark{b}\tablefootmark{c}} & B1Ib & 4.91 & 24000 & 16.5 & 10 & 36\tablefootmark{a} & $ 2.5\times10^{35 }$ &$ 1.4\times10^{-8  }$ & 0.04 & 67, 68 \\
\object{2S 0114+650\tablefootmark{b}} & B1Iae & 5.61 & 24000 & 37 & 16 & 56 & $ 1.1\times10^{36 }$ &$ 1.6\times10^{-7  }$ & 0.24 & 1, 2, 37 \\
\object{Vela X-1\tablefootmark{b}} & B0Ia & 5.49 & 25500 & 28.4 & 20.2 & 50 & $ 3.5\times10^{36 }$ &$ 1.3\times10^{-7  }$ & 0.19 & 5, 56, 69 \\
\object{IGR J11215-5952\tablefootmark{c}} & B0.5Ia & 5.73 & 24700 & 40 & 29 & 80\tablefootmark{a} & $ 3\times10^{36 }$ &$ 2.7\times10^{-7  }$ & 0.35 & 19, 20 \\
\object{1E 1145.1-6141\tablefootmark{b}} & B2Iae & 5.12 & 19500 & 32 & 14 & 63 & $ 1.1\times10^{36 }$ &$ 3.9\times10^{-8  }$ & 0.11 & 70, 71 \\
\object{GX 301-2\tablefootmark{b}} & B1.5Iae & 5.67 & 18100 & 70 & 43 & 180 & $ 2\times10^{37 }$ &$ 4.7\times10^{-7  }$ & 0.73 & 72, 73, 74 \\
\object{OAO 1657-415} & B3Ia & 4.95 & 20000 & 24.8 & 14.3 & 50.3 & $ 3\times10^{36 }$ &$ 1.8\times10^{-8  }$ & 0.05 & 75, 76 \\
\object{IGR J18029-2016} & B1Ib & 5.14 & 25000 & 19.8 & 20.2 & 33.1 & $ 2\times10^{36 }$ &$ 3.7\times10^{-8  }$ & 0.05 & 10, 29 \\
\object{IGR J18483-0311\tablefootmark{c}} & B0.5Ia & 5.57 & 24600 & 33.8 & 33 & 96 & $ 3.7\times10^{35 }$ &$ 1.6\times10^{-7  }$ & 0.24 & 14, 15, 42 \\
\hline
\end{tabular}
\tablefoot{Stellar parameters were taken from the listed references, except for
the mass-loss rates, for which we used the fits from \citet{bcmfkont},
neglecting clumping,
and for the optical depth 
parameters, which were
calculated from Eq.~\eqref{tx} using the fits for
the
mass-loss rates and terminal velocities from \citet{bcmfkont}.
\tablefoottext{a}{Periastron distance.}
\tablefoottext{b}{Some alternative designations: 
IGR J00370+6122 (BD+6073),
2S 0114+650 (V662 Cas),
Vela X-1 (GP Vel, HD 77581),
1E 1145.1-6141 (V830 Cen),
and  GX 301-2 (BP Cru).}
\tablefoottext{c}{Supergiant fast X-ray transient \citep{lutov,walt,hmxb57}.}
}
\tablebib{(1)~\citet{hmxb1};
(2)~\citet{hmxb2};
(5)~\citet{viteal};
(10)~\citet{hmxb10};
(14)~\citet{hmxb14};
(15)~\citet{hmxb15};
(19)~\citet{hmxb19};
(20)~\citet{hmxb20};
(29)~\citet{hmxb29};
(37)~\citet{hmxb37};
(42)~\citet{hmxb42};
(56)~\citet{hmxb56};
(67)~\citet{hmxb67};
(68)~\citet{hmxb68};
(69)~\citet{sandvelax};
(70)~\citet{hmxb70};
(71)~\citet{hmxb71};
(72)~\citet{hmxb72};
(73)~\citet{hmxb73};
(74)~\citet{hmxb74};
(75)~\citet{hmxb75};
(76)~\citet{hmxb76}.}
\end{table*}

In more realistic models that account for clumping, the mass-loss rate is higher
and the recombination becomes stronger. Therefore, the zone of the wind
inhibition recedes in the X-ray luminosity versus optical depth diagram (see
Fig.~\ref{xiobrch}). This alleviates the problem of binaries that appear in the
region of wind inhibition. Moreover, a significant fraction of binary systems
appears close to the inhibition region boundary, which is what indicates that their X-ray
luminosities may be self-regulated \citep{irchuch}.

Out of the studied models, the coolest one (150-60) is least influenced by X-rays.
This is caused by the combination of its highest mass-loss rate, slow velocity
leading to a denser wind \citep{vinbisja}, higher X-ray opacity due to neutral
helium, and generally weaker ionization with no irradiation.

Many of the HMXB primaries listed in Table~\ref{neutron} have radii comparable
to the orbital separation. If they evolve toward red parts of the
Hertzsprung-Russell diagram, then their radius quickly approaches the binary
separation, as inferred from evolutionary models of \citet{sylsit}. This puts
the purely wind accretion commencing Roche lobe overflow phase \citep{tuty} to
its end. This offers an explanation for the lack of HMXBs with late-supergiant
primaries \citep{lipar} provided that the binary separation does not
significantly expand over the course of evolution \citep{dvojvyjvit}.

\section{Predicting the X-ray luminosity}

The X-ray emission of wind-powered HMXBs originates due to the release of
gravitational potential energy during accretion of the stellar wind
\citep{davos,laheupet}. The resulting X-ray luminosity is affected by processes
acting on very different spatial scales, from the scales comparable to binary
separation \citep{manousci,xustone}, at which the global properties of the flow
are determined, across the magnetospheric radius of the neutron star
\citep{nerozumim,bozof}, which determines the way in which the material enters the
magnetosphere (if there is a magnetosphere), down to scales comparable to
a neutron star radius or Schwarzschild radius.

The accretion rate and associated accretion luminosity can be determined within
the approximate Bondi-Hoyle-Lyttleton theory \citep{holy,boho}, which despite
its numerous simplifications provides reasonable estimates for the accretion rate in
many circumstances \citep{xustone}. With $\x M$ and $\x R$ denoting the mass and
radius of an accreting object, respectively, and $v$ being its velocity relative to
the wind flow, the Bondi-Hoyle-Lyttleton accretion luminosity is
\citep{laheupet}
\begin{equation}
\label{lxlxrov}
\lx=\frac{G^3\x M^3}{\x R D^2 v^4}\dot M.
\end{equation}
Here $G$ is the gravitational constant. This equation gives the maximum X-ray
luminosity, which assumes a maximum efficiency for the Bondi-Hoyle-Lyttleton accretion
mechanism \citep[e.g.,][]{martpreh,sipo}. The relative velocity can be estimated
using the orbital velocity of the compact component $v_\mathrm{orb}$ and the
wind velocity at the distance $D$ of the compact component, $v_\mathrm{wind} =
v(D),$ as
\begin{equation}
\label{pytha}
v^2=v_\mathrm{wind}^2+v_\mathrm{orb}^2.
\end{equation}
For simplicity, we inserted $v_\mathrm{wind} = v_\infty$.

\begin{figure}
\includegraphics[width=0.43\textwidth]{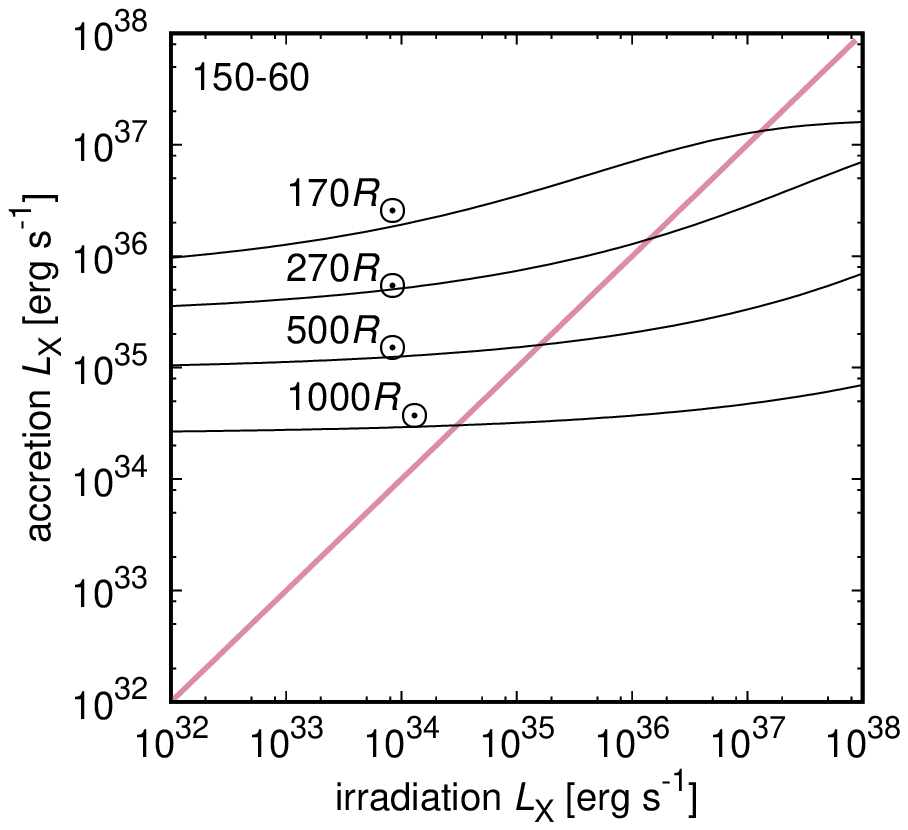}
\includegraphics[width=0.43\textwidth]{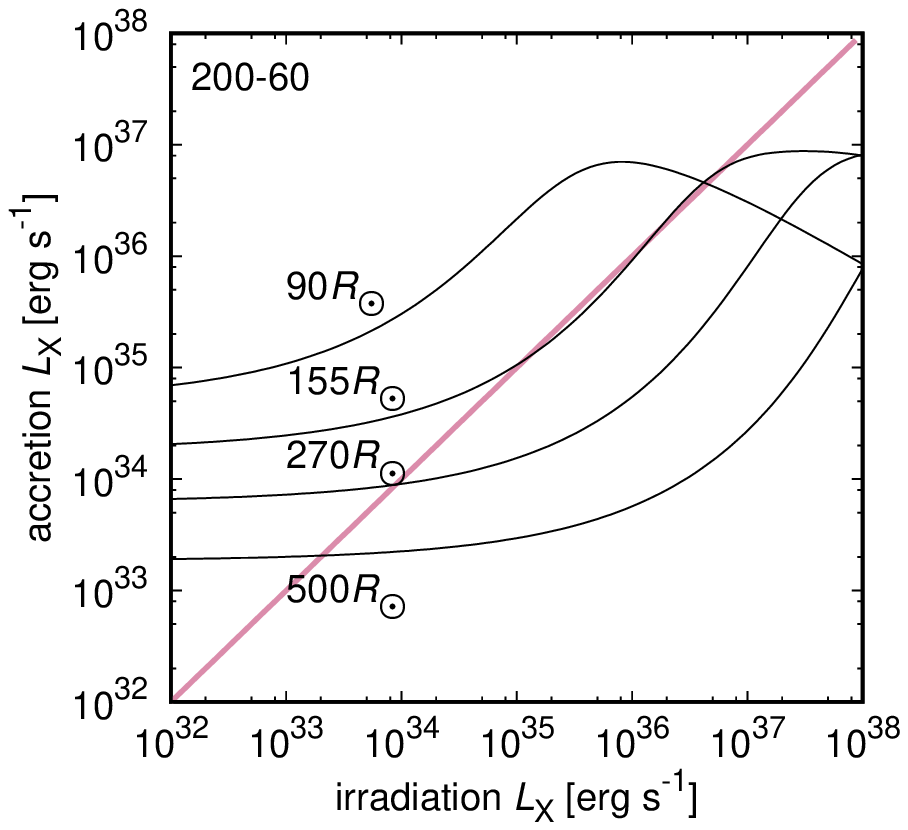}
\includegraphics[width=0.43\textwidth]{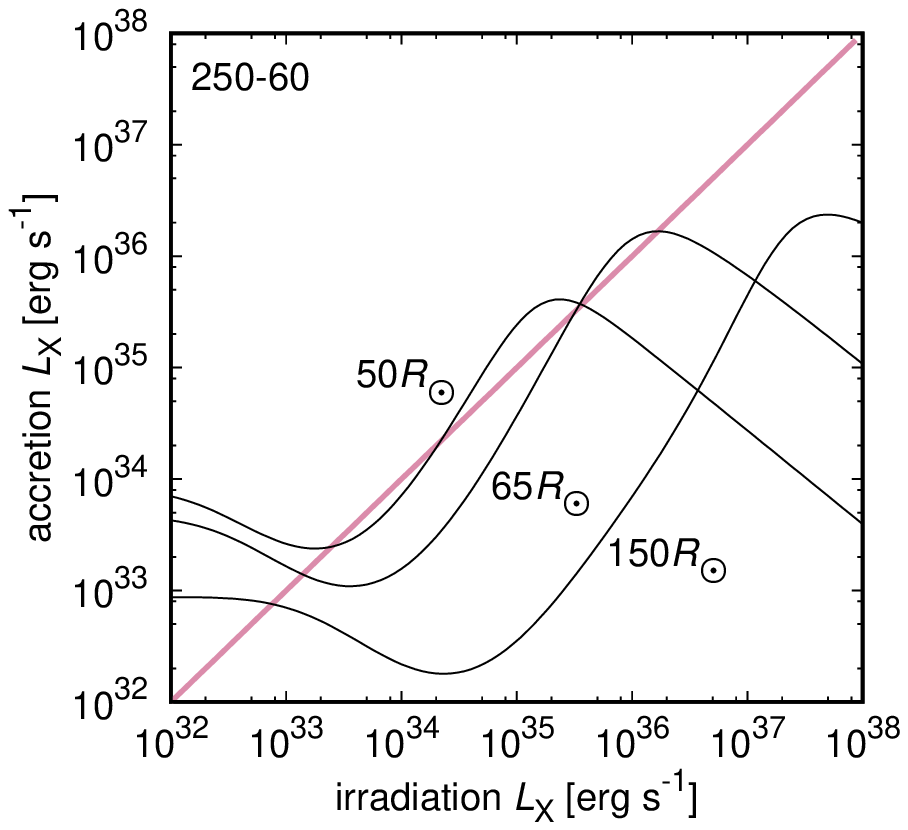}
\centering
\caption{X-ray luminosity generated by wind accretion as a function of X-ray
irradiation after Eq.~\eqref{lxlxrov} accounting for the dependence of the
terminal velocity and the mass flux on X-ray irradiation via
Eqs.~\eqref{ch10najventaurov} and \eqref{ch10najventaurovdmdt}. 
The figure was plotted for
individual model stars from Table~\ref{bvele} and for different binary
separations labeled in the plots. The pink line denotes the one-to-one relation.}
\label{lxlxlx}
\end{figure}

For the given stellar and binary parameters, Eq.~\eqref{lxlxrov} provides the X-ray
luminosity due to wind accretion as a function of the wind velocity and mass
flux. However, these wind parameters are modified by X-ray irradiation
(Eqs.~\eqref{ch10najventaurov} and \eqref{ch10najventaurovdmdt}). Consequently,
Eq.~\eqref{lxlxrov} predicts the accretion X-ray luminosity as a function of X-ray
irradiation. This function is plotted in Fig.~\ref{lxlxlx} for individual model
stars and for different orbital separations. For the plots we assumed generic
neutron star parameters $\x M=1.4\,M_\odot$ and $\x R=10\,\text{km}$.

For weak X-ray irradiation, the wind terminal velocity and the mass flux remain
unaffected. Wind flows at large velocities; therefore, only a small fraction of
the wind is collected by the compact companion, and consequently the resulting
X-ray emission is weak (Fig.~\ref{lxlxlx}). For stronger X-ray irradiation, the
terminal velocity decreases; as a result, the compact companion accretes more wind
and the X-ray emission becomes much stronger \citep{hohoho,karbim}. When the
wind velocity is negligible with respect to the orbital velocity, $v\approx
v_\text{orb}=\sqrt{GM/D}$ from Eq.~\eqref{pytha}, the X-ray luminosity reaches
its maximum value, which is $\lx=G\x M\dot M(\x
M/M)^2/\x R$ from Eq.~\eqref{lxlxrov}, that is to say it is a factor of $(M/\x M)^2$ lower than one would get from
a complete accretion of the stellar wind. We note that the maximum X-ray
luminosity does not depend on the binary separation; therefore, even binaries
with relatively large separations may have strong X-ray luminosities. However,
even the mass flux decreases for very large X-ray luminosities. This leads to
inhibition of the wind and X-ray emission (Fig.~\ref{lxlxlx}). For lower binary
separations, the influence of X-rays becomes stronger; therefore, the plots shift
to the left and their maxima shift to lower values of $\x L$ in
Fig.~\ref{lxlxlx}.

In a stationary state, the irradiation X-ray luminosity is equal to the
accretion luminosity \citep{karbim,irchuch,boznebim}. Consequently,
Eq.~\eqref{lxlxrov} can be regarded as an implicit equation for \lx. Individual
solutions of this equation correspond to the points where the curves of X-ray
luminosity intersect with the one-to-one relation in Fig.~\ref{lxlxlx}. For large
binary separations, there is just one root of Eq.~\eqref{lxlxrov} corresponding
to low X-ray luminosity and a weak influence of X-rays. For medium separation,
there may be up to three solutions corresponding to different X-ray luminosities
and different strengths of the influence of X-rays. The solutions with low X-ray
luminosities disappear for small binary separations and only a solution with a high
luminosity and strong influence of X-rays remains.

Not all of discussed solutions are stable. If the slope of the function is
steeper than the slope of the one-to-one relation, then a small perturbation leads
to runaway from the initial solution \citep{irchuch}. From the lower plot of
Fig.~\ref{lxlxlx}, it follows that this happens for the middle solutions.
Therefore, only solutions with the largest and the lowest X-ray luminosities are
stable.

We applied Eq.~\eqref{lxlxrov} to determine the X-ray luminosities of HMXBs with
{\Bsupergiant} components listed in Table~\ref{neutron}. We inserted their
stellar and binary parameters and accounted for the influence of X-rays on their
terminal velocities (via Eq.~\eqref{ch10najventaurov}). Most stars show a solution
with a strong decrease in the wind velocity and high X-ray luminosity on the order
of $10^{36}\,\ergs$. This value corresponds to typical values found from
observations (column $\x L$ in Table~\ref{neutron}).

A class of HMXBs is characterized with relatively low quiescent X-ray
luminosities $10^{32}-10^{33}\,\ergs$ and sporadic outbursts reaching X-ray
luminosities on the order of $10^{36}\,\ergs$. These binaries are called
supergiant fast X-ray transients (SFXTs) and objects that belong to this group
are marked in Table~\ref{neutron} by a superscript $c$. \citet{karbim} and
\citet{irchuch} propose that low quiescent X-ray luminosities of SFXTs
correspond to low-luminosity solutions of Eq.~\eqref{lxlxrov}, for which the
wind is not significantly affected by X-rays. \citet{boznebim} argue that the
transitions from the low luminosity state to the high luminosity state and back
are modulated by changing the binary separation on highly eccentric orbits. Here
we additionally demonstrate a bimodality of wind solutions in {\Bsupergiant}
HMXBs, which appears for particular system parameters due to stronger
sensitivity of the wind velocity on X-ray luminosity (higher $\beta_1$ and
$\beta_2$ in Eq.~\eqref{ch10najventaurov}), and which is absent in
O\nobreakdash-supergiant HMXBs \citep{irchuch}.

In addition to the solutions with a high luminosity, some of the SFXTs in Table~\ref{neutron} also
have low-luminosity solutions. This might be an explanation for their SFXT
properties, but the distance variations during orbital motion may also be
important, as suggested by \citet{boznebim}.

From Fig.~\ref{lxlxlx}, it follows that only binaries with a relatively small
separation $D\lesssim200\, R_\odot$ are strong X-ray sources with X-ray
luminosities on the order of $10^{36}\,\ergs$. This means that there can be a
large population of X-ray quiet binaries with degenerate companions and X-ray
luminosities on the order of $10^{33}\,\ergs$. This is a typical X-ray
luminosity of single OB stars \citep{igor}; therefore, these binaries may be
hidden among B supergiants without compact companions.

\begin{figure}[t]
\centering
\resizebox{\hsize}{!}{\includegraphics{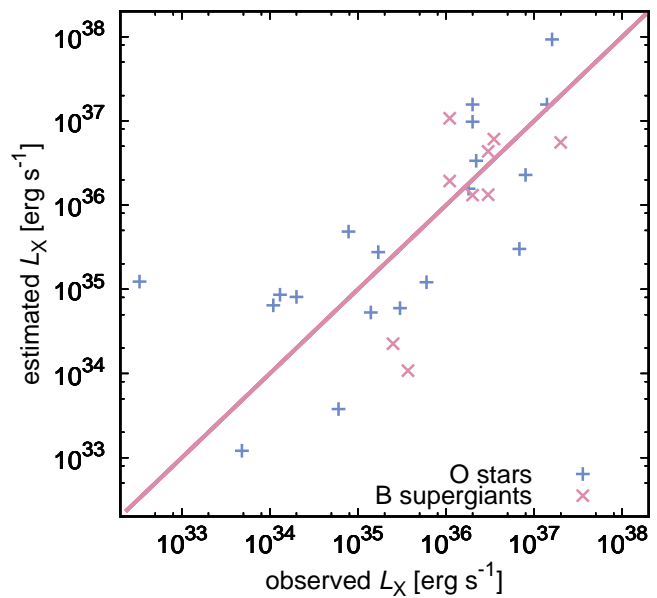}}
\caption{X-ray luminosities of {\Bsupergiant} HMXBs (from Table~\ref{neutron})
and O\protect\nobreakdash-star HMXBs \citep[from the list of][]{irchuch}
estimated using Eqs.~\eqref{lxlxrov} and \eqref{pytha} inserting the wind
terminal velocity derived from Eq.~\eqref{ch10najventaurov} calculated using
observed X-ray luminosity. Plotted against observed X-ray luminosity. The pink line
denotes the one-to-one relation.}
\label{lxlxv}
\end{figure}

It is possible to combine our sample of {\Bsupergiant} HMXBs with
O\nobreakdash-star HMXBs from \citet{irchuch} and to solve Eq.~\eqref{lxlxrov}
to derive an estimate for the X-ray luminosity of these binaries. While this
approach provides a good estimate for the X-ray luminosity of stars with high
observed luminosities $\lx>10^{36}\,\ergs$, predictions for binaries with low
X-ray luminosities are typically overestimated. This perhaps happens because the
curve predicted by Eq.~\eqref{lxlxrov} is closely aligned with the one-to-one
relation and; therefore, a small change in the binary parameters may cause a large
change in the predicted X-ray luminosity.

To alleviate this problem, we inserted the wind terminal velocity determined via
Eq.~\eqref{ch10najventaurov} using observed X-ray luminosity into
Eq.~\eqref{lxlxrov}. The predicted X-ray luminosities estimated in this way
nicely agree with observed values (see Fig.~\ref{lxlxv}). A good agreement
between observed and estimated X-ray luminosities results from a decrease in the
wind velocity due to X-ray irradiation \citep{sandvelax,irchuch}. Without taking
the influence of X-rays on the wind velocity into account, the estimated values
of $\x L$ are by one to two orders of magnitude lower.

\begin{figure}
\includegraphics[width=0.5\textwidth]{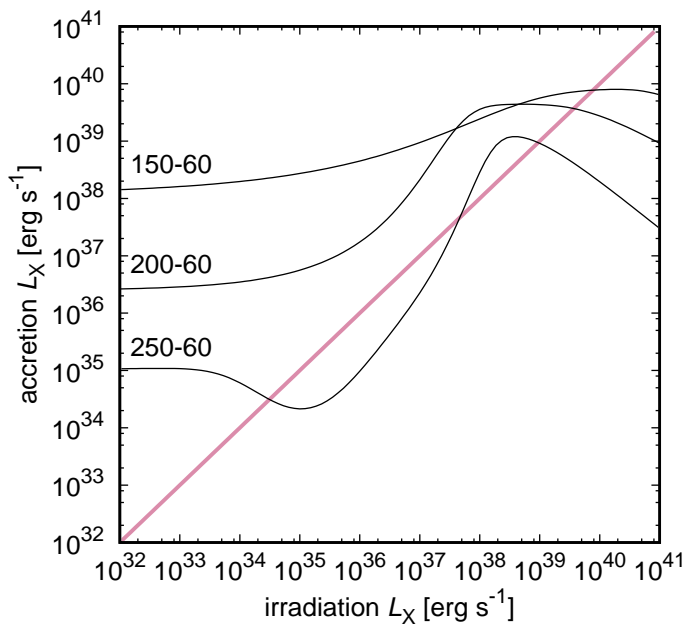}
\centering
\caption{X-ray luminosity generated by wind accretion on a $20\,M_\odot$ black
hole located at $D=300\,R_\odot$ as a function of X-ray irradiation after
Eq.~\eqref{lxlxrov}. Plotted for individual model stars from Table~\ref{bvele}.
The pink line denotes the one-to-one relation.}
\label{cdlxlxlx}
\end{figure}

\begin{figure*}
\includegraphics[width=\textwidth]{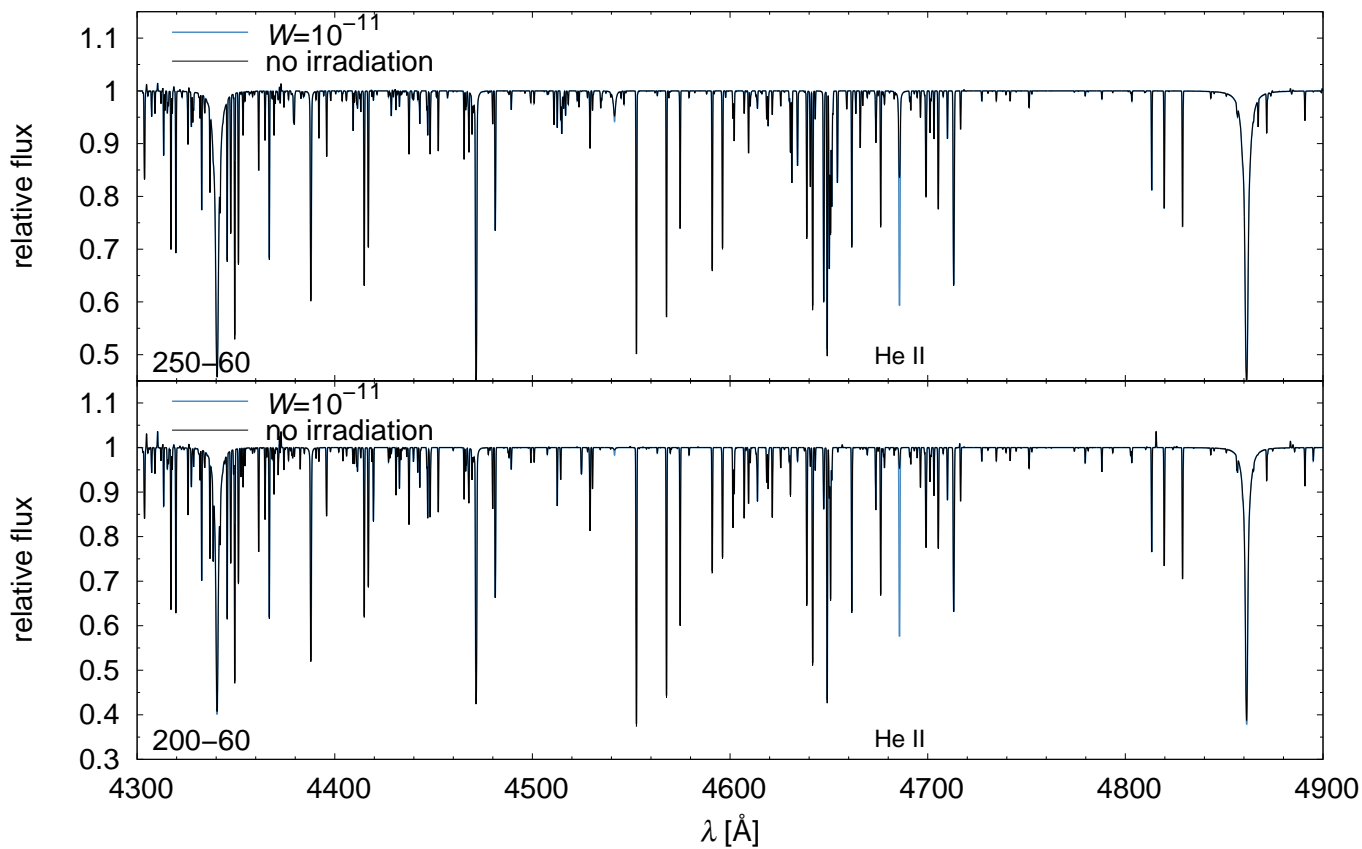}
\caption{Comparison of spectra of selected model stars with and without X-ray
irradiation.}
\label{bveloz}
\end{figure*}

\section{Implications for ULXs}

Typical ULXs have X-ray luminosities in excess of $10^{39}\,\ergs$
\citep{cernyulx,waltulx}. Although the ULXs are believed to be powered mostly by
Roche lobe overflow \citep{karulx,elmelulx}, the wind accretion remains a viable
option at least for some of these sources \citep{milmakaron,wiktor}.

From Fig.~\ref{lxlxlx} it follows that the maximum X-ray luminosity stemming
from the accretion of a {\Bsupergiant} wind on a neutron star is on the order of
$10^{37}\,\ergs$. Therefore, a higher mass for the compact object is required to
obtain luminosities in the ULX regime. In Fig.~\ref{cdlxlxlx} we plotted the
accretion X-ray luminosity as a function of the X-ray irradiation for the binaries
where the compact companion is a black hole with a mass of $20\,M_\odot$ instead
of a neutron star. The value for the black-hole mass is motivated by an updated
mass of Cygnus X-1 derived from radio interferometry \citep{cygx1bum}. As a
result of cubic dependence of the X-ray luminosity on the compact-companion
mass, the resulting X-ray luminosities are by two to three orders of magnitude
higher than those for a neutron star. For B supergiants below the bistability
jump, the maximum possible X-ray luminosities correspond to the weaker end of
ULX regime.

X-ray luminosities corresponding to a weak effect of X-ray irradiation are by
several orders of magnitude lower (Fig.~\ref{cdlxlxlx}). Therefore, there might
exist a population of quiet HMXBs with massive black holes as a central engine
and X-ray luminosities on the order of $10^{35}-10^{36}\,\ergs$.

Strong X-ray irradiation not only inhibits wind acceleration, but it may also
affect the spectrum of B supergiants. To understand this effect, we included the
X-ray irradiation in TLUSTY atmosphere models \citep{ostar2003,bstar2006}. The
external irradiation is included as an outer boundary condition for a specific
intensity,
\begin{equation}
I_\nu^\text{irrad}=W B_\nu(T_\irr),
\end{equation}
where $W=(1/4)(R_\irr/D)^2$ is a dilution factor, $R_\irr$ is the effective
radius of the irradiating body, and $B_\nu(T_\irr)$ is Planck law at the
temperature $T_\irr$. With $\x L=4\pi R_\irr^2 \sigma T_\irr^4$, the dilution
factor $W$ is related to the X-ray luminosity and source distance as
\begin{equation} W=\frac{\x L}{16\pi D^2\sigma T_\irr^4}=7\times10^{-12}
\zav{\frac{\x L}{10^{40}\,\text{erg}\,\text{s}^{-1}}}
\zav{\frac{D}{100\,R_\odot}}^{-2}. \end{equation} Taking the possible
parameters of ULXs into account, we calculated two sets of atmosphere models with
$W=10^{-11}$ and $W=10^{-13}$ and with $T_\irr=10^7\,$K.

Even for strong X-ray irradiation, the resulting spectra are nearly
indistinguishable from the spectra without any irradiation (Fig.~\ref{bveloz}).
Although the X-rays are able to heat the outer regions of the photosphere by several
tens of thousands Kelvin, only layers with a low Rosseland optical depth
$\tau_\text{ross}\lesssim10^{-3}$ are affected, and therefore the influence of X-ray
irradiation on emergent optical spectrum is relatively weak. As a result, only a
few absorption lines (e.g., \ion{He}{ii} 4686\,\AA, 5412\,\AA, and 6560\,\AA)
become stronger in the irradiated spectra due to an increase in populations of
excited levels of \ion{He}{ii} by incoming X-rays. There are additional lines in
the infrared domain which show enhanced emission due to irradiation, for
instance the \ion{He}{i} 18\,685\,\AA\ line. Irradiated spectra given in
Fig.~\ref{bveloz} correspond to the maximum irradiation in the surface region
that directly faces the neutron star. Other regions receive less flux,
and consequently the integrated effect is expected to be even lower.

\section{Conclusions}

We studied the effect of X-ray irradiation on the stellar wind in HMXBs powered
by the accretion of {\Bsupergiant} wind on its compact companion. We included an
external X-ray source in our {\Bsupergiant} wind models. For each model star
corresponding to B supergiants, we calculated a grid of models parameterized by
the binary separation and external irradiation X-ray luminosity. We also
calculated models with optically thin inhomogeneities (clumping).

It is well known that accretion-generated X-rays alter the ionization state of
the wind. Higher ionization states, which appear due to X-ray ionization, drive
the wind less effectively and, consequently, brake acceleration of the wind.
This causes a decrease in the wind terminal velocity and, for strong X-ray
irradiation, also a decrease in the mass flux in the direction of the companion. These effects
are particularly important for short binary separations and high X-ray
luminosities. The X-ray ionization can be partially compensated for by wind
clumping, which increases recombination and mass-loss rates. The influence of
clumping is particularly strong in the region of the bistability jump,
where the mass-loss rate increases toward lower effective temperatures by a factor
of a few as a result of iron recombination.

The strength of X-ray illumination can be conveniently demonstrated in the
diagrams that plot (undisturbed) optical depth parameter versus the X-ray
luminosity. There is a parameter region of high X-ray luminosities and low
optical depth parameters in these diagrams, where the X-ray ionization leads to
the disruption of the wind. Observational parameters of high-mass X-ray binaries
with B supergiant components appear outside the zone of wind disruption, which is in
agreement with clumped-wind model results. Moreover, a significant fraction of
HMBXs appears close to the border of wind disruption indicating that their X-ray
luminosities may be self-regulated.

The X-ray feedback determines the X-ray luminosity resulting from wind
accretion. We recognized two states of feedback. For low X-ray luminosities,
the X-ray ionization is weak, the wind is not disrupted by X-rays, it flows at
large velocities, and consequently the accretion rate is relatively low. On the
other hand, for high X-ray luminosities, the X-ray ionization disrupts the flow
braking the acceleration of the flow facing the companion, the wind velocity is
low, and the accretion rate becomes high. We demonstrated that these effects
determine the X-ray luminosity of individual binaries. By accounting for wind
inhibition by X-rays, the estimated X-ray luminosities are consistent with
observational values. Moreover, the two states of X-ray feedback can explain the
appearance of two types of X-ray binaries, classical supergiant X-ray binaries
and fast X-ray transients.

In HMXBs with massive black hole components, the X-ray luminosities may exceed
$10^{39}\,\ergs$ for B supergiants below the bistability jump. This shows that
part of the ULXs may be powered by the accretion of {\Bsupergiant} wind. Despite
the presence of a strong illuminating source, the optical spectrum of such supergiants
is nearly indistinguishable from the spectrum of single B supergiants.

\begin{acknowledgements}
Computational resources were supplied by the project "e-Infrastruktura CZ"
(e\nobreakdash-INFRA LM2018140) provided within the program Projects of Large
Research, Development and Innovations Infrastructures. The Astronomical
Institute Ond\v{r}ejov is supported by a project RVO:67985815 of the Academy of
Sciences of the Czech Republic.
\end{acknowledgements}


\bibliographystyle{aa}
\bibliography{papers}

\end{document}